\documentclass[noshowpacs,prl,aps,superscriptaddress,floatfix,twocolumn]{revtex4-2}
\usepackage[normalem]{ulem}
\usepackage{amsmath}
\usepackage{graphicx}
\usepackage{hyperref}
\hypersetup{
     colorlinks=true,
     linkcolor=blue,
     filecolor=blue,
     citecolor = orange,      
   }
\usepackage[dvipsnames]{xcolor}
\usepackage{siunitx}
\def\sss{\scriptscriptstyle}
\begin{document}
\title{
Topological Pathways to Two-Dimensional Quantum Turbulence }
\author{R. Panico}
\affiliation{CNR NANOTEC, Institute of Nanotechnology, Via Monteroni, 73100 Lecce, Italy}
\affiliation{Institut für Angewandte Physik, Universität Bonn, Wegelerstraße 8, 53115 Bonn, Germany}
\author{G. Ciliberto}
\affiliation{Universit\'e Paris-Saclay, CNRS, LPTMS, 91405, Orsay, France}
\author{G. I. Martone}
\affiliation{CNR NANOTEC, Institute of Nanotechnology, Via Monteroni, 73100 Lecce, Italy}
\affiliation{INFN, Sezione di Lecce, 73100 Lecce, Italy}
\author{T. Congy}
\affiliation{Department of Mathematics, Physics and Electrical Engineering, Northumbria University, Newcastle upon Tyne NE1 8ST, United Kingdom}
\author{D. Ballarini}
\affiliation{CNR NANOTEC, Institute of Nanotechnology, Via Monteroni, 73100 Lecce, Italy}
\author{A. S. Lanotte} 
\affiliation{CNR NANOTEC, Institute of Nanotechnology, Via Monteroni, 73100 Lecce, Italy}
\affiliation{INFN, Sezione di Lecce, 73100 Lecce, Italy}
\author{N. Pavloff}
\affiliation{Universit\'e Paris-Saclay, CNRS, LPTMS, 91405, Orsay, France}
\affiliation{Institut Universitaire de France (IUF)}

\begin{abstract}
  We present a combined experimental and theoretical investigation of
  the formation and decay kinetics of vortices in two-dimensional,
  compressible quantum turbulence. We follow the temporal evolution of
  a quantum fluid of exciton–polaritons, hybrid light–matter
  quasiparticles, and measure both phase and modulus of the order
  parameter in the turbulent regime. Fundamental topological conservation laws
  require that the formation and annihilation of vortices
  also involve critical points of the velocity field, namely nodes and
  saddles.
  Identifying the simplest mechanisms underlying these processes
  enables us to develop an effective kinetic model that closely aligns
  with the experimental observations, and shows that
  different processes are responsible for vortex number
  growth and decay.
  These findings underscore the crucial role played by topological
  constraints in shaping nonlinear, turbulent evolution of
  two-dimensional quantum fluids.
\end{abstract}

\maketitle

Topological and dynamical properties of two-dimensional systems are
strongly intertwined. This is true not only in condensed matter setups
\cite{Qi2011,Culcer2020,Reichhardt2022} but also for hydrodynamical
systems, be these classical or quantum. In classical fluids the
identification of topological critical points proves helpful for
classifying flow patterns \cite{Lighthill1966,Perry1975} and studying
two-dimensional spatio-temporal chaos and turbulence
\cite{Moffatt2001,Rossi2006,Ouellette2008,Garcia2014,Smith2017}.  As
for quantum fluids, the importance of quantization of vorticity has
been understood long ago \cite{Onsager1949,Feynman1955} and vortices
indeed play a major role in the route to two-dimensional quantum
turbulence
\cite{Nazarenko2006,Neely2013,White2014,Sachkou2019,Gauthier2019,Johnstone2019,Forrester2020,Eloy2021,Abobaker2023,Panico2023},
as they do in the classical context
\cite{Mcwilliams_1984,Babiano1987,Benzi1987,Brachet1988,Boffetta2012}.
In this Letter, we further explore the link between dynamical and
topological properties in two-dimensional quantum turbulence. We
propose to investigate the temporal properties of the quantum fluid
velocity field by a novel strategy. The idea is to devise a minimal model 
which complies with global topological
constraints, without requiring local knowledge of the spatial dynamics
of the system.  To achieve this, we derive kinetic equations of
formation and annihilation of critical points of the velocity field,
and apply the approach to a non-equilibrium exciton-polariton
fluid. We show that we can reproduce the experimentally observed rate
of creation and annihilation of quantized vortices, thus identifying
the elementary mechanisms responsible for the increase in the number
of vortices --during the quantum turbulence growth-- and for its
reduction --during the quantum turbulence decay.

We consider a two-dimensional quantum fluid described by a scalar
order parameter of the form
$\psi (\vec{r}, t)=A(\vec{r}, t) \exp\{ i \Theta (\vec{r}, t)\}$. Here the
real functions $A$ ($\geq 0$) and $\Theta$ correspond to the amplitude and
phase of the order parameter, respectively, and $\vec{r} = (x,y)$. The
velocity field of the fluid is $\vec{v}=(\hbar/m) \vec{\nabla} \Theta$
\cite{Feynman1955}. In a two-dimensional setting, two topological
indices are associated with any domain $D$ delimited by a close
contour $C$, namely, the vorticity $I_{\rm V}$ and the Poincar\'e
index $I_{\rm P}$ \cite{Nye1988}
\begin{equation}
      I_{\rm V}=\frac{1}{2\pi} 
  \oint_C {\rm d}\Theta, \;\;\;\;
I_{\rm P}=\frac{1}{2\pi}\oint_C
    {\rm d}\varphi,
\end{equation}
where $\varphi$ denotes the polar angle of $\vec{v}$.
\begin{figure}
\includegraphics[width=\linewidth]{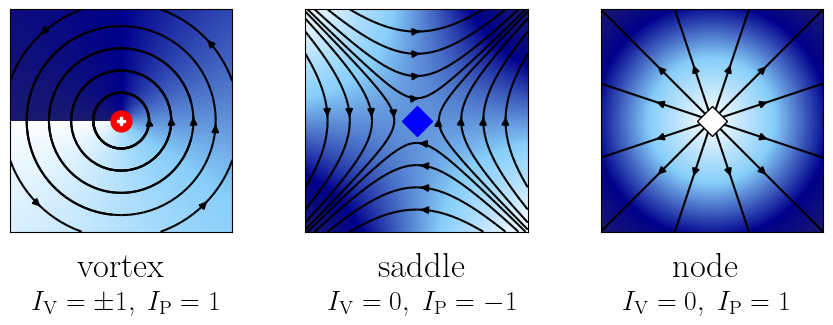}
\caption{Sketch of the streamline pattern around a (positive) vortex,
  a saddle, and a node (phase minimum). Darker regions correspond to
  larger values of the phase $\Theta$ of the order parameter. A vortex
  is a branch point of the phase, the corresponding branch cut is
  represented by a discontinuity of the color map in the left
  plot. Nodes and saddles are stagnation points where
  $\vec{v}=\vec{0}$.}
\label{fig:cp}
\end{figure}
$I_{\rm V}$ is (up to a factor $2\pi$) the variation of the phase
$\Theta$ along the contour $C$. $I_{\rm P}$ is the net algrebraic
number of revolutions made by the velocity field's direction along $C$
\cite{strogatz_nonlinear_2015}. It is interesting to note that what is
commonly referred to as the vorticity in the context of the two
dimensional xy model is actually the Poincar\'e index; see, e.g.,
\cite{Kosterlitz_1973,Kosterlitz_1974}.  Both indices are zero if
there are no singular nor stagnation points inside $D$.  They assume
nontrivial values when the phase $\Theta$ displays extrema (local
maxima or minima), saddles, or essential singularities. The
corresponding points are nodes (attractive or repulsive), saddles, and
quantum vortices, respectively.  Figure~\ref{fig:cp} gives the values
of the indices attached to each of these points, which we loosely
denote as critical points in the following.  The vorticity and
Poincar\'e index attached to a given domain are the sum of the indices
of all the critical points it contains.

\begin{figure}
\includegraphics[width=\linewidth]{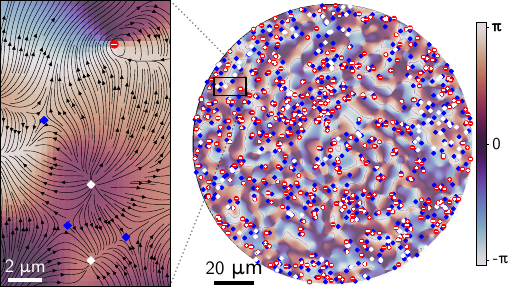}
\caption{A snapshot of the polariton superfluid phase field, with the
  measured critical points. The three types of critical points are
  represented with the same symbols as in Fig. \ref{fig:cp}.  A zoom,
  with streamlines represented as oriented solid lines, highlights the
  local flow organization, revealing three saddles, two nodes (one
  attractive and one repulsive), and a negative vortex.}
\label{fig:0}
\end{figure}

The co-existence of the three types of critical points presented in
Fig. \ref{fig:cp} has been explicitly experimentally demonstrated in
linear \cite{Shvartsman1995} and nonlinear \cite{Congy2024} optics.
The physical system we examine here involves injecting a high-energy
polariton superfluid, and allowing it to expand within a circular
potential barrier \cite{Panico2023}. The initial kinetic energy
provided to the superfluid induces the creation not only of a dense
vortex gas but also of a large number of saddles and nodes.  The
optical nature of polaritons allows for the measurement of both the
modulus and the phase of the order parameter through interferometric
techniques \cite{Caputo2019,Sitnik2022,Abobaker2023}, which enables
recording the flow pattern with a level of detail currently
unattainable in other types of superfluids.  As shown in
Fig. \ref{fig:0}, by analyzing the velocity field, we can track the
evolution of hundreds of critical points.
This method enables us to determine, at each time step, the number of
vortices, saddles, and nodes present in the system. We emphasize that
the presence of nodes is a unique feature of compressible and
non-stationary quantum fluids, such as polariton superfluids. These
nodes are indeed observed in our experimental results, and we
demonstrate below that they play a crucial role in the onset of
turbulence.

{\it Model - }We consider the following main mechanisms of creation
(or annihilation) of critical points in the flow field: (i) the
nodes-to-vortices conversion in which two nodes coalesce and give
birth to two vortices and (ii) the saddle-node bifurcation which
creates one saddle and one node from scratch. These two processes 
conserve the vorticity and the Poincar\'e index; they correspond to
well-identified bifurcations whose relevance for a two-dimensional
quantum fluid has been validated in Ref. \cite{Congy2024}. They can be
schematically written as chemical reactions:
\begin{subequations}\label{eq1}
\begin{align}
\mbox{node}\, + \, \mbox{node}\;  & 
\overset{a}{\underset{b}{\rightleftharpoons}}
\;\mbox{vortex}_{(+)}\, + \, \mbox{vortex}_{(-)}\, \label{eq1a} \; , \\
  \O \;\;
& \overset{c}{\underset{d}{\rightleftharpoons}}
\;\mbox{node}\, + \, \mbox{saddle} \; ,\label{eq1b}
\end{align}
\end{subequations}
where vortices with positive or negative vorticity are denoted as
vortex$_{(+)}$ or vortex$_{(-)}$, respectively. The (positive)
quantities $a$, $b$, $c$, and $d$ are the reaction rates, see
Eq. \eqref{prey_predator1} below.  Mechanism \eqref{eq1a} appeared
implicitly in works by Indebetouw \cite{Indebetouw1993} and the Soskin
group \cite{Soskin1997}, then explicitly in
Ref. \cite{Soskin1999}. Mechanism \eqref{eq1b} is mentioned by Freund
in Ref. \cite{Freund1999}.  Other mechanisms have been observed
\cite{Congy2024} which also conserve both the vorticity and the
Poincar\'e index: a saddle can transform into two saddles plus one
node in a pitchfork bifurcation, or also a vortex-antivortex pair and
two saddles can appear spontaneously (or coalesce) in a process first
identified by Nye, Hajnal, and Hannay \cite{Nye1988} which has been
termed the ``Bristol mechanism'' in Ref. \cite{Congy2024}. These
reactions have been discarded for simplicity reasons (they involve
collisions of a larger number of critical points) and also because
much less often observed in a previous experiment and in numerical
simulations \cite{Congy2024}.

\begin{figure}
\includegraphics[width=1\linewidth]{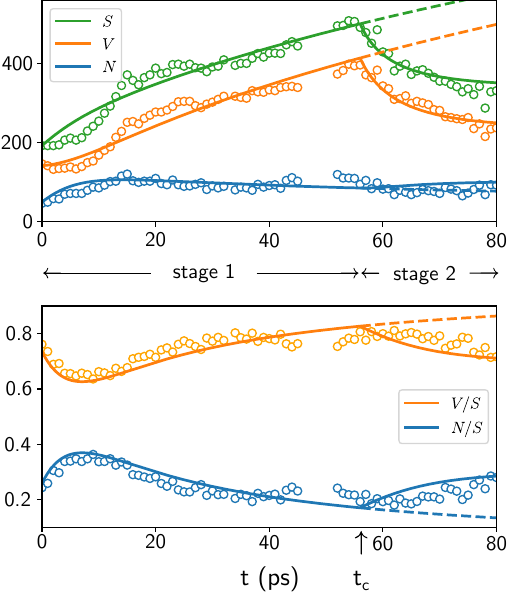}
\caption{(Top) Comparison of the experimental results for $N(t)$,
  $V(t)$ and $S(t)$ (circles) with the theoretical predictions
  (lines). Experimental data are averages of four realisations of the
  same dynamical regime. For $t\le t_{\rm c}$, the solid lines have
  been obtained with the numerical integration of
  Eqs. \eqref{eq.rescaled} with the values $\gamma=0.52$,
  $N_0=170$, and $t_0=11$ ps \cite{footnote_caption}. For $t> t_{\rm c}$,
  the dashed lines correspond to the results of \eqref{eq.rescaled},
  while the solid lines come from the numerical resolution of
  \eqref{eq.rescaled2} with $\varepsilon = 0.045$. (Bottom)
  Same as above for the quantities $V(t)/S(t)$ and $N(t)/S(t)$. The
  value of $t_{\rm c}$ is $56$ ps.}
\label{fig.high_energy}
\end{figure}

From the modeling \eqref{eq1}, we write a kinetic equation
inspired by rate equations of elementary chemical reactions: 
\begin{equation}\label{prey_predator1}
\begin{split}
& \dfrac{{\rm d} V_\pm}{{\rm d}t} = a N^2 - b V_+ V_- ,
\qquad \dfrac{{\rm d}S}{{\rm d}t} = c - d N S ,
\\
& \dfrac{{\rm d}N}{{\rm d}t} = - 2 a N^2 + 2 b V_+ V_- +c -d N S,
\end{split}
\end{equation}
where $N(t)$ denotes the number of nodes, $S(t)$ the number of
saddles, and $V_{+}(t)$ [$V_{-}(t)$] the number of vortices with
positive [negative] vorticity. It results from the values of the
topological indices listed in Fig. \ref{fig:cp} that the total
Poincar\'e index of the system is
$I_{\rm\sss P} = N + V_+ + V_- - S$. It is easily verified that
$I_{\rm\sss P}$ is preserved by the system \eqref{prey_predator1}:
this comes as no surprise since the elementary processes \eqref{eq1}
both conserve the Poincar\'e index.  Similarly, the conserved
total vorticity of the system is $V_+-V_-$.  In the following we make
the simplifying assumption that this difference is equal to zero:
$V_+(t) = V_-(t) = V(t)/2$ where $V(t)$ is the total number of
vortices. This hypothesis is confirmed by the experimental data (such
as displayed in Fig. \ref{fig.high_energy}) and is certainly sound in
the configuration we consider where typically $V(t)\gg 1$ while no
external angular momentum is imparted to the system.

Defining the rescaled quantities $\tau=t/t_0$, $n = N/N_0$, $v=V/N_0$,
and $s=S/N_0$, with $t_0=1/\sqrt{2ac}$ and $N_0=\sqrt{c/2a}$, makes it
possible to cast the system \eqref{prey_predator1} under the following
dimensionless form:
\begin{equation}\label{eq.rescaled}
\begin{split}
& \dfrac{{\rm d}v}{{\rm d}\tau} = n^2-\alpha v^2, 
\qquad
\dfrac{{\rm d}s}{{\rm d}\tau} = 1 - \gamma ns,\\
& \dfrac{{\rm d}n}{{\rm d}\tau} = 1 - n^2 - \gamma ns +\alpha v^2,
\end{split}
\end{equation}
where $\alpha=b/(4a)$ and $\gamma=d/(2a)$
\footnote{For completeness, note that in terms of the new parameters,
the rates of reaction read $a=1/2 N_0 t_0$, $b=2 \alpha/N_0 t_0$, $c=N_0/t_0$,
and $d=\gamma/N_0 t_0$.}.

{\it Results - }We consider a turbulent regime of the polariton
dynamics in which, after fast expansion of the quantum fluid, the
onset of vortex clustering and the emergence of the inverse kinetic
energy cascade was evidenced on timescales of a few tens of
picoseconds \cite{Panico2023}. The numbers of vortices, saddles and
nodes, extracted from the data of Ref. \cite{Panico2023}, are
displayed as circles in the top part of Fig. \ref{fig.high_energy}. At
$t=0$, when the fluid hits the barrier, some critical points are
already present, having formed during the fluid's expansion. The
turbulent dynamics is initiated at this moment, which we treat as the
initial condition. A low-energy data set, where the onset of
turbulence is inhibited by dissipation, is presented in
\cite{supplemental} along with additional details on the experimental
configuration.  Let us first focus on the stage of turbulence growth,
during which the numbers of vortices and saddles increase
significantly (stage 1 in Fig.~\ref{fig.high_energy}). In this time
lag, the nucleation of many new vortices and saddles dominates the
temporal evolution. This implies imposing $\alpha=0$: indeed, when
$\alpha\ne 0$ the system \eqref{eq.rescaled} has a fixed point and the
numbers of vortices, saddles, and nodes tend to saturate, which is not
what is observed in the experiment.  We checked that a nonzero value
of $\alpha$ always worsens the agreement of the theoretical curve with
data: this confirms that in this stage the incompressible kinetic
energy of the system is mostly increasing, as required for the
establishment of the inverse cascade of kinetic energy, see the
Discussion section below.

It is interesting to discuss the values of the rate of reactions in
Eqs. \eqref{eq1}. In particular $c/d=N_0^2/\gamma=6\times 10^4\gg 1$,
implying that the saddle-node bifurcation is mainly unidirectional:
the annihilation of a saddle with a node is much less frequent than
their {\it creatio ex nihilo}.  This indicates that the saddle-node
formation mechanism \eqref{eq1b} is the real fuel of the whole
process. The nodes-to-vortices reaction \eqref{eq1a} merely transmutes
some of the nodes into vortices, but could not be effective on its
own.  This remark is of significance: the spontaneous creation of
uniquely a vortex-antivortex pair being topologically forbidden (it
would not conserve the Poincar\'e index) we are in need of an
explanation of the increase of the number $V(t)$ of vortices. In the
system we consider, the formation of vortices arises from two
saddle-nodes bifurcations \eqref{eq1a} followed by a nodes-to-vortices
conversion \eqref{eq1b}, ultimately resulting in the formation of two
saddles and two vortices. This is the reason why, as shown in the top
part of Fig. \ref{fig.high_energy}, the numbers of saddles and of
vortices increase at the same pace. The results plotted in the bottom
panel of Fig. \ref{fig.high_energy} indicate that the total Poincar\'e
index is conserved and small. Indeed in this case $N+V=S$, the two
quantities $V/S$ and $N/S$ sum to unity, and a minimum of one should
correspond to a maximum of the other. This property is
model-independent: it is a prerequisite which should be embodied in
any kinetic model, but its fulfillment is not a guarantee of accuracy
of the model. Experimental results confirm the exact conservation of
both $I_{\rm V}$ and $I_{\rm P}$ indices in every realization of the
measurements.

The results displayed in Fig. \ref{fig.high_energy} show a striking
behavior, namely, a sharp temporal transition from stage 1,
characterised by the nonlinear growth of the number of
vortices and saddles, to stage 2, characterised by a dramatic decrease of
the number of vortices and
saddles.
However, the number of nodes is not experiencing a similar abrupt
modification in the same period of time: this supports a scenario
which does not involve nodes, still conserving both $I_{\rm V}$ and
$I_{\rm P}$.
The so-called Bristol mechanism \cite{Nye1988}, described by Eq. 
\eqref{eq.Bristol} below, is a perfect candidate:
\begin{equation}\label{eq.Bristol}
  \mbox{vortex}_{(+)}\, + \,
  \mbox{vortex}_{(-)}\, + \, \mbox{saddle} \, + \,
  \mbox{saddle} \; 
  \overset{e}{\underset{f}{\rightleftharpoons}} \; \; \O
\end{equation}

In view of the significant decrease of the number of vortices and
saddles during stage 2, we consider that the rate of reaction $f$ is
zero in Eq. \eqref{eq.Bristol}. Hence, the process is assumed to be
unidirectional \footnote{This unidirectional behavior was already
observed in the experiment and the numerical simulations of
\cite{Congy2024}: In this reference the Bristol mechanism was always
inducing the concomitant annihilation of two vortices and two saddles,
and never their {\it creatio ex nihilo}.}. The system
\eqref{eq.rescaled} accordingly modifies to
\begin{equation}\label{eq.rescaled2}
\begin{split}
& \dfrac{{\rm d}v}{{\rm d}\tau} = n^2-\alpha v^2
-\varepsilon\, v^2 s^2, \quad \dfrac{{\rm d}s}{{\rm d}\tau} = 1 - \gamma ns
-\varepsilon\, v^2 s^2,\\
& \dfrac{{\rm d}n}{{\rm d}\tau} = 1 - n^2 - \gamma ns +\alpha v^2,
\end{split}
\end{equation}
where $\varepsilon=\tfrac12 e N_0^3 t_0=e c/(8 a^2)$ is the rescaled
rate of annihilation of saddles and vortices. We keep for all the
other parameters the values previously determined, and during stage 2
we
solve the system \eqref{eq.rescaled2} with $\varepsilon\neq 0$. The
corresponding results are displayed in Fig. \ref{fig.high_energy}. The
agreement of the theoretical curve with the experimental observation
supports the idea that after $t=t_c$ the system enters a new
regime in which the annihilation mechanism \eqref{eq.Bristol} acquires
an efficiency it previously did not have.

It is interesting to ask the question whether the mechanism of
Eq. \eqref{eq.Bristol} --which is explicitly observed in our
experiment \cite{supplemental}-- could have been effective earlier,
with a rate of reaction $f\ne 0$ explaining the rapid and concomitant
increase of $V$ and $S$ during stage 1. The observation of the
behavior of $N$ in the same period makes this hypothesis rather
unlikely, since $N$ initially increases and then saturates. This
advocates for a saddle-node creation process \eqref{eq1b} which then
feeds the nodes-to-vortices one \eqref{eq1a}. Only this process can
explain (i) the occurrence of extrema of $V/S$ and $N/S$ at short
times (bottom plot of Fig. \ref{fig.high_energy}) and (ii) the
saturation of $N$ at a slightly later time (top plot of the same
figure). And indeed, it is not possible to accurately reproduce the
experimental data on the basis of mechanisms \eqref{eq1a} and
\eqref{eq.Bristol} only, or \eqref{eq1b} and \eqref{eq.Bristol} only.

{\it Discussion - } The sharp modification of the time evolution of
the number of vortices and saddles at $t_c=56$ ps is well described by
the inclusion of the Bristol mechanism \eqref{eq.Bristol}, but the
very fact that such a transition occurs is not explained by our
model. We show here that this transition occurs exactly at the time
where the inverse turbulent cascade stops.

\begin{figure}
\includegraphics[width=\linewidth]{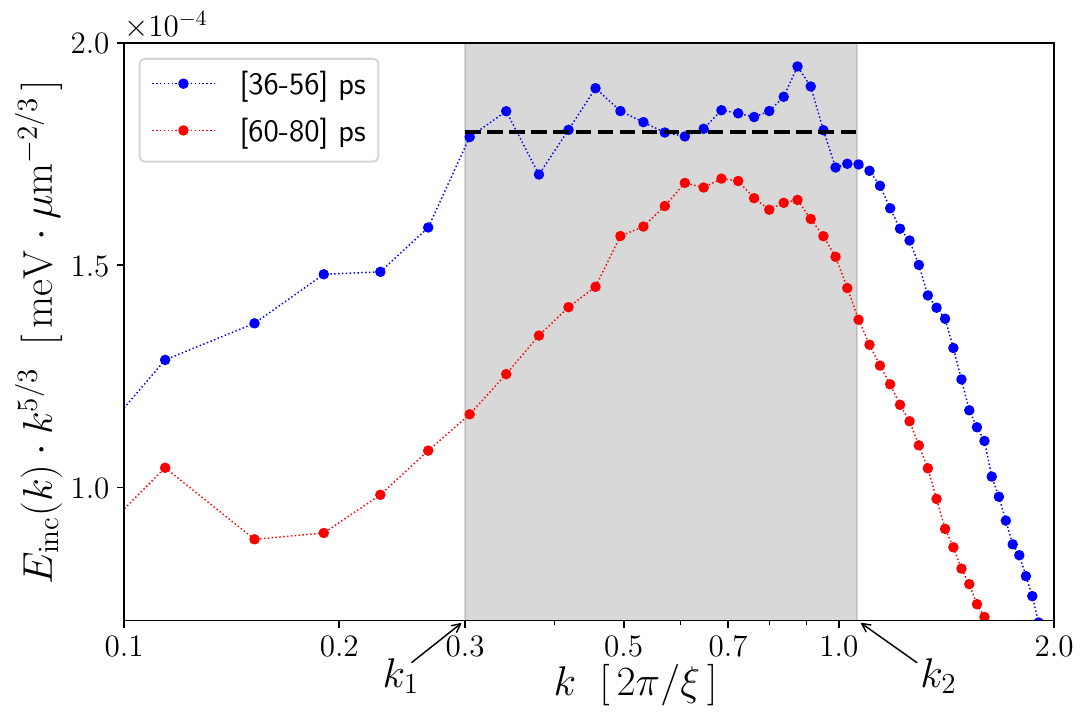}
\caption{Compensated experimental spectra of the incompressible
  kinetic energy, averaged over two different time windows: during the
  stage in which the inverse cascade develops ([36-56] ps), and during
  the decay stage ([60-80] ps). The grey area identifies the spectral
  region of wavenumbers, the so-called {\it inertial range},
  associated to the inverse cascade of the incompressible kinetic
  energy $k_{1} < k <k_{2}$, where $k_{1}\xi/2\pi=0.3$ and
  $k_{2}\xi/2\pi=1.06$, $\xi=\hbar (2 m g |\psi|^2)^{-1/2}$ being the
  healing length. The horizontal dashed line is just a guide for the
  eye.}
\label{fig:spectra}
\end{figure}

\begin{figure}
\includegraphics[width=\linewidth]{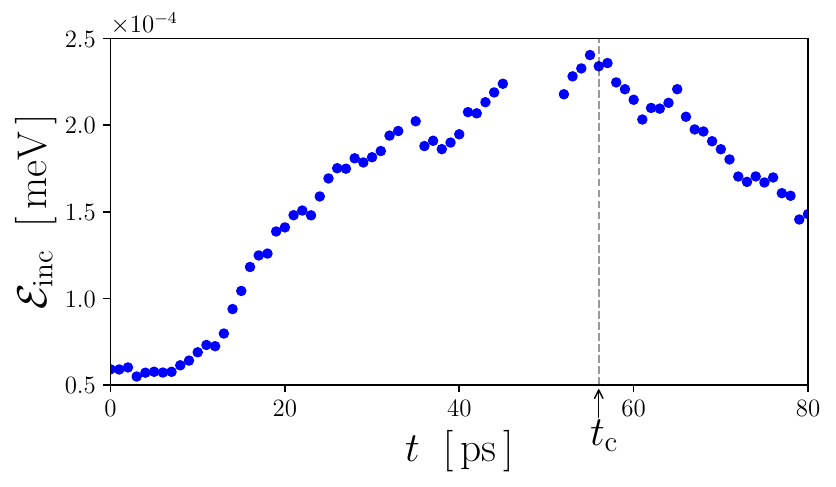}
\caption{Time
  evolution of the incompressible kinetic energy ${\cal E}_{\rm inc}(t)$
  defined in Eq. \eqref{eq.Einc}.}
\label{fig:energy}
\end{figure}

In Fig. \ref{fig:spectra} we re-analyse the data of
Ref. \cite{Panico2023} by displaying the experimental one-dimensional
spectra of the incompressible kinetic energy $E_{\rm inc}(k)$, where
$k= |\vec{k}|$ \footnote{The density-weighted superfluid velocity
field $\vec{u}(\vec{r},t)=|\psi|\vec{v}$, where $|\psi|^2$ is the
number density of the polaritons, is separated into two components: a
divergence-free one ($\vec{u}_{\rm inc}$), which is the incompressible
part, and an irrotationnal one ($\vec{u}_{\rm comp}$), which is the
compressible part. The one-dimensional spectral density of the
incompressible kinetic energy is obtained by integrating over the
polar angle: $E_{\rm inc}(k,t)= \tfrac12 m \, k\, \int d\theta_k \,
|\vec{u}_{\rm inc}(\vec{k},t)|^2$, see Ref. \cite{Nore1997} and the
supplementary information of Ref. \cite{Panico2023}.}, averaged over
two different time windows. In the grey area for wavenumbers $k_1 < k
<k_2$, the average of spectra measured for time lags $t\in [36,56]$ ps
(blue points in Fig. \ref{fig:spectra}) exhibits a behavior compatible
with the expected Kolmogorov-like scaling \cite{Kraichnan1967},
$E_{\rm inc}(k) \propto k^{-5/3}$ \footnote{A Kolmogorov-like scaling
for the inverse energy cascade in compressible flows has been reported
for classical fluids in Ref. \cite{Puggioni2020}. Its experimental
realisation in compressible quantum flows \cite{Panico2023} is a
non-trivial observation by itself.}. This tendency no longer persists
beyond $t_c$: the average of spectra measured for $t\in [60,80]$ ps
(red points in Fig. \ref{fig:spectra}) displays a narrower scaling
region and a smaller amplitude. These are both indications of the end
of the inverse cascade.
This trend is further confirmed by an analysis of the temporal
behaviour of the part of the incompressible kinetic energy contained
within the inertial range of the inverse cascade (i.e., for
$k\in[k_1,k_2]$), that we call ${\cal E}_{\rm inc}(t)$. This quantity
is defined by
\begin{equation}\label{eq.Einc}
  {\cal E}_{\rm inc}(t) \equiv \int_{k_1}^{k_2}\!\! E_{\rm inc}(k,t) \,dk.
\end{equation}
Its evaluation is made possible by the recording at each time lag of
the experimental spectrum $E_{\rm inc}(k,t)$.  ${\cal E}_{\rm
  inc}(t)$, plotted in Fig. \ref{fig:energy}, is an estimate of the
energy available to establish the inverse cascade process. The onset
of a turbulent inverse cascade of kinetic energy implies a temporal
growth of the incompressible part of the total kinetic energy in the
system. Indeed, the results show that, after set-up time, ${\cal
  E}_{\rm inc}(t)$ goes on growing as expected, until the critical
time $t_c=56$ ps. At this stage the available incompressible kinetic
energy starts its decay and can no longer sustain the inverse transfer
process across scales.

The fact that the crossover time $t_c$ is observed during the growth
then simultaneous rapid decrease of both the vortices and saddles
numbers, and that $t_c$ also marks the end of the temporal growth of
the incompressible kinetic energy, suggests that these processes are
fed by the incompressible kinetic energy available in the inertial
range.  When this stops growing, dissipation mechanisms prevail and
turbulence starts decaying. Interestingly, at the same time the
clustering dynamics stops \cite{supplemental}.

{\it Conclusion - }In the experiments we presented, dynamical
observations associated to the turbulence growth/decay have their
topological counterpart in the time window where the numbers of
vortices and saddles increase/decrease. It is reasonable to think that
not all vortices participate in the cascade, since they may not have
time to correlate, nevertheless their increase reflects in the growth
of the incompressible kinetic energy available for the
cascade. \\ Topological constraints also rule the mechanism of the
turbulence decay; a process based on four-vortex
interactions \footnote{Note that since in the observed dynamics,
saddles and vortices have similar temporal evolutions, the decay due
to the Bristol mechanism \eqref{eq.Bristol} is effectively equivalent
to a four-vortex decay process.} previously proposed in
\cite{Nazarenko2007,Groszek2016,Karl2017,Baggaley2018,Kanai2024}, here
finds its origin in topological arguments. In the absence of a
turbulent regime, the fate of vortices is different. In such a case,
we physically expect a dynamical equilibrium between vortex creation
and annihilation processes, in the presence of random, uncorrelated
fluctuations. Our model faithfully describe this process, see
\cite{supplemental}.

The kinetic model here introduced is the simplest that complies with
topological constraints. It provides a global, averaged description of
the system based on phenomenological parameters (the rate
coefficients) but is not designed to explain why these parameters
assume different values in the turbulent or non-turbulent regimes, nor
to predict when turbulence growth halts and why its decay is so
abrupt. Addressing these phenomena requires to account for vortex
clustering, i.e., to deal with spatial correlations within the system.

This focus on spatial correlations is crucial in the study of
two-dimensional turbulence: since Polyakov’s pioneering contribution
\cite{Polyakov1993} it has been shown that the vorticity domains
exhibit the same universal scaling arising in critical percolation
theory, in both classical \cite{Bernard2006,Puggioni2020} and quantum
\cite{Panico2023b} fluids in the regime of inverse energy cascade.
Broadening the scope of our kinetic approach to set up a microscopic
model that integrates these statistical properties would therefore be
of great interest.  Such a model should account for interactions
between critical points (such as vortex clustering) within a framework
consistent with the conservation of topological indices.

\

{\it Acknowledgments - } N. P. acknowledges insightful comments by
M. V. Berry and M. R. Dennis.  D. B., T. C. and A. L. acknowledge the
kind hospitality during the workshop ``Turbulence and Vortex dynamics
in 2D quantum fluids'' at the International Center for Theoretical
Sciences (ICTS, Bangalore, India), where part of this work was
discussed (code: ICTS/QUFLU2024/2). T. C. and N. P. thank the Isaac
Newton Institute for Mathematical Sciences, Cambridge, for support and
hospitality during the programme Emergent Phenomena in Nonlinear
Dispersive Waves, where the work on this paper was partially
undertaken. This work was supported by the Italian Ministry of
University and Research (MUR) through the PNRR MUR project: `National
Quantum Science and Technology Institute' - NQSTI (PE0000023) and the
PNRR MUR project: `Integrated Infrastructure Initiative in Photonic
and Quantum Sciences' - I-PHOQS (IR0000016). We acknowledge the
support of the Quantum Optical Networks based on Exciton-polaritons -
(Q-ONE) funding from the HORIZON-EIC-2022-PATHFINDER CHALLENGES EU
programme under grant agreement No. 101115575, and of the Neuromorphic
Polariton Accelerator - (PolArt) funding from the
Horizon-EIC-2023-Pathfinder Open EU programme under grant agreement
No. 101130304. Views and opinions expressed are however those of the
author(s) only and do not necessarily reflect those of the European
Union or European Innovation Council and SMEs Executive Agency
(EISMEA). Neither the European Union nor the granting authority can be
held responsible for them.

{\it Data availability -} The data that support the findings of this
article are available upon reasonable request from the authors.

\bibliography{biblio}

\begin{thebibliography}{58}
\bibitem [{foo()}]{footnote_caption}%
  \BibitemOpen
  \href@noop {} {}\bibinfo {note} {The quality of the agreement does not
  sensibly depend on the values of the parameters $\gamma$, $N_0$, and $t_0$.
  Numerical checks indicate that the values of these fitting parameters are
  fixed with a relative uncertainty of the order of 10 \%.}\BibitemShut {Stop}%
\end{thebibliography}

\end{document}